\documentclass{appolb}
\usepackage{epsfig}
\title{RESUMMATION OF DOUBLE LOGARITHMIC TERMS $LN^2(1/x)$ IN THE POLARIZED
 NONSINGLET STRUCTURE FUNCTION $g_1$ AT SMALL $x$ VIA GLAP-LIKE APPROACH.}

\author{Dorota Kotlorz $^1$, Andrzej Kotlorz $^2$
\address{$^1$Department of Physics Ozimska 75, $^2$Department of 
Mathematics Luboszycka 3, Technical University of Opole, 
45-370 Opole, Poland, e-mail $^1$: {\tt dstrozik@po.opole.pl}}}

\begin{document}
\pagestyle{plain}
\eqsec
\maketitle

\begin{abstract}
An alternative equation, resumming of the $\ln^2 1/x$ terms for the
polarized nonsinglet structure function $g_1^{NS}$ at small $x$ is
presented. Construction of the GLAP-like formula for the auxiliary function,
corresponding to the $g_1^{NS}$ at rescaled $Q^2$ variable is shown.
Predictions of this approach for the $g_1^{NS}$ function at small $x$ in
a case of a flat as well as a dynamical input are given. The role of the
fixed coupling constant and the running one is also discussed.
\end{abstract}


\section{Introduction}

Our knowledge about the structure functions of the nucleon is still
incomplete because of lack of understanding how these structure functions
behave in the small Bjorken $x$ region. Neither present experimental data
nor the theoretical QCD description give a full and unique picture of an
exact shape of the quark and gluon distributions in the nucleon.
Perturbative QCD analysis, based on the GLAP evolution equations \cite{b1} 
is in a good agreement with experimental measurements. This agreement concerns 
unpolarized \cite{b2}, \cite{b3} and polarized \cite{b4} structure functions of 
the nucleon within NLO approximation in the large and moderatory small Bjorken 
$x$ region. Unfortunately, practically lack of experimental data in the small 
$x$ region ($x<10^{-3}$) makes the satisfactory determination of the Bjorken 
and Ellis-Jaffe sum rules \cite{b5} impossible. This causes e.g. 
that the question "how is the spin of the nucleon made of partons?" remains 
still open. From recent papers \cite{b6}, \cite{b7} we know, that the small $x$
behaviour of both unpolarized and polarized structure functions is
controlled by the double logarithmic terms $\alpha_s\ln^21/x$. However, this
singular behaviour of the structure functions at low $x$ is better visible
in the spin dependent case. For the unpolarized, nonsinglet structure
function $F_2^{NS}(x,Q^2)=F_2^p(x,Q^2)-F_2^n(x,Q^2)$ the QCD singular
behaviour at small $x$ is overridden by the leading Regge contribution
\cite{b8}. Next in the unpolarized, singlet case, the structure function at 
low $x$ is driven by BFKL pomeron \cite{b9} because gluons play the dominant 
role. Thus the growth of structure functions of the nucleon, governed by
leading double logarithmic terms $\alpha_s^n\ln^{2n}(x)$ becomes best
visible for spin dependent functions. Therefore the polarized structure
functions of the nucleon may be a sensitive test of the perturbative QCD
analyses in the small $x$ region. The double logarithmic $\ln^2x$ effects go
beyond the standard LO or even NLO QCD evolution of parton distributions and
correspond to the ladder diagrams with quark and gluon exchanges along the
ladder. One has also to take into account nonladder diagrams but in the
nonsinglet case they may be neglected as nonleading \cite{b6}, \cite{b7}. Thus 
the nonsinglet, polarized structure function
$g_1^{NS}(x,Q^2)=g_1^{p}(x,Q^2)-g_1^{n}(x,Q^2)$ is a convenient function
both for QCD analyses (because of its simplicity) and future experimental
tests at HERA \cite{b10} concerning the determination of the Bjorken sum rule.
Theoretical predictions for $g_1^{NS}$ at small $x$, incorporating the
double $\ln^2x$ effects have been presented in \cite{b7}, \cite{b11}, 
\cite{b12}, \cite{b13}. In these papers the perturbative QCD analysis is based 
on the unintegrated spin dependent quark distributions $f(x,k^2)$. It means 
that the sum of double logarithmic $\ln^2x$ terms is represented by a 
appropriate integral equation for the unintegrated structure function 
$f(x,k^2)$.

In this paper we present an alternative approach of the double $\ln^2x$
resummation for $g_1^{NS}$ at small $x$. Our formalism is based on the usual
quark distribution functions (and not on $f(x,k^2)$ function) and moreover
generates the $\ln^2x$ terms via GLAP-like evolution equation for the
rescaled quark transverse momentum squared $\mu^2=k^2/x$. The purpose of
this paper is to compare these two methods of the double $\ln^2x$
resummation for the polarized nonsinglet structure function $g_1^{NS}$ at
low $x$. In the next section we briefly recall the origin of the double
logarithmic $\ln^2x$ effects at low $x$, incorporating the evolution
equation based on the unintegrated function $f(x,k^2)$. In section 3 we
introduce alternative formalism, in which after rescaling the kinematic
variable $\mu^2=k^2/x$ we get GLAP-like evolution equation in $\mu^2$. This
equation for the polarized quark distributions (and hence for the $g_1^{NS}$
function) generates the double logarithmic $\ln^2x$ terms. Section 4
contains numerical results for the spin dependent nonsinglet structure
function $g_1^{NS}(x,Q^2)$ in our approach. We compare the both mentioned
above methods and also compare their predictions for $g_1^{NS}$ with SMC
1997 small $x$ data. Finally in section 5 we summarize our results.

\section{Double logarithmic $\ln^2x$ resummation for the nonsinglet,
polarized structure function $g_1^{NS}(x,Q^2)$ using the unintegrated
function $f(x,k^2)$.}

It has been lately noticed \cite{b6}, \cite{b7} that the spin dependent 
structure function $g_1$ in the small $x$ region is dominated by $\ln^2(1/x)$ 
terms. This singular behaviour, implied by QCD is for the polarized structure 
functions the leading one. Comparatively, for unpolarized, nonsinglet structure
functions of the nucleon, the QCD evolution behaviour at small $x$ is
screened by the leading Regge contribution. The Regge theory \cite{b14}, which
concerns the Regge limit: $x\rightarrow 0$ predicts the following behaviour
of parton distributions at small $x$ and $Q^2\le 1~{\rm GeV}^2$:
\begin{eqnarray}\label{r2.1}
x\Sigma&\sim& const~(Pomeron)\nonumber\\
q_{NS}&\sim& x^{-0.5}~(Reggeon~A_2:\rho -\omega);\nonumber\\
\Delta\Sigma,\Delta q_{NS}&\sim& x^0\div x^{0.5}~(Reggeon~A_1)
\end{eqnarray}
where $\Sigma$, $q_{NS}$, $\Delta\Sigma$, $\Delta q_{NS}$ denote
respectively singlet unpolarized, nonsinglet unpolarized, singlet polarized,
nonsinglet polarized quark distributions. The shape of all spin dependent
distributions $\Delta\Sigma$, $\Delta q_{NS}$ is mostly governed by QCD
evolutions with dominating $\ln^2x$ terms at small $x$. These $\ln^2x$
contributions correspond to the ladder diagram with quark and gluon
exchanges along the ladder - {\it cf} Fig.1.
\begin{figure}[ht]
\begin{center}
\includegraphics[width=80mm]{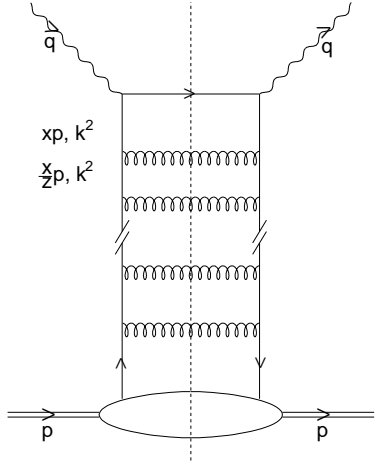}
\caption{A ladder diagram generating double logarithmic $\ln^2(1/x)$
terms in the nonsinglet spin structure function $g_1$.}
\end{center}
\end{figure}
In contrast to the singlet polarized function, for the nonsinglet one the
contribution of nonladder diagrams is negligible. Thus examing the
polarized, nonsinglet structure function $g_1^{NS}(x,Q^2)$, we should
consider only mentioned above ladder diagrams. The nonsinglet part of the
spin dependent structure function has a form:
\begin{equation}\label{r2.2}
 g_1^{NS}(x,Q^2)=g_1^p(x,Q^2)-g_1^n(x,Q^2)
\end{equation}
where $g_1^p$ and $g_1^n$ are spin dependent structure functions of
proton and neutron respectively. Let us remind the meaning of $g_1$.
In the Bjorken limit
\begin{equation}\label{r2.3}
g_1(x)=\frac12 \sum\limits_{i=u,d,s,..} e_i^2 \Delta q_i(x)
\end{equation}
\begin{equation}\label{r2.4}
\Delta q_i(x)=q_{i+}(x)-q_{i-}(x)
\end{equation}
where $e_i$ is a charge of the i-flavour quark, $q_{i+}(x)$
$(q_{i-}(x))$ is the density distribution function of the
i-quark with the spin parallel (antiparallel) to the parent nucleon.
Function $g_1(x,Q^2)$ is connected with the helicity of the nucleon
({\it i.e.} spin projection on the momentum direction). Thus the integral
\begin{equation}\label{r2.5}
\langle\Delta q_i\rangle =\int\limits_0^1 \Delta q_i(x) dx
\end{equation}
is simply a part of the nucleon helicity, carried by a quark of
i-flavour (i=u,d,s,..). Polarized distribution functions of quarks are defined as:
\begin{equation}\label{r2.6}
\Delta q=\Delta q_{val}+\Delta q_{sea}
\end{equation}
Finally:
\begin{equation}\label{r2.7}
g_1^{NS}=\frac16(\Delta u_{val}-\Delta d_{val})=\frac16(\Delta u-\Delta d)
\end{equation}
Solutions of the equation for the unintegrated polarized nonsinglet structure
function $f^{NS}(x,Q^2)$, which gives the $\ln^2x$ resummation are presented
in \cite{b7}, \cite{b11}, \cite{b12}, \cite{b13}. $\ln^2x$ as the only 
contribution at small $x$  behaviour of $g_1^{NS}$ is examined in \cite{b7} 
while in \cite{b11}, \cite{b12}, \cite{b13} the unified description of 
$g_1^{NS}$ incorporating both Gribov-Lipatov-Altarelli-Parisi (GLAP) evolution
and $\ln^2x$ effects is presented. These all approaches are based on the 
unintegrated distribution function $f(x,Q^2)$, which is related to the 
$g_1(x,Q^2)$ via
\begin{equation}\label{r2.8}
g_1(x,Q^2)=g_1^{(0)}(x)+\int\limits_{k_0^2}^{Q^2(1/x-1)}
\frac{dk^2}{k^2}f\left(x(1+\frac{k^2}{Q^2}),k^2\right)
\end{equation}
where
\begin{equation}\label{r2.9}
g_1^{(0)}(x)=\int\limits_0^{k_0^2}\frac{dk^2}{k^2}f(x,k^2)
\end{equation}
Let us recall from \cite{b11} that the resummation of the double logarithmic
terms $\ln^2x$ in the limit of a very small $x$ ($x\rightarrow 0$) is given
in the case of the polarized, nonsinglet structure function by:
\begin{equation}\label{r2.10}
f_{NS}(x,k^2)=f_{NS}^{(0)}(x,k^2)+\frac{\alpha_s(k^2)}{2\pi}
\int\limits_x^1\frac{dz}{z}\int\limits_{k_0^2}^{k^2/z}
\frac{dk'^2}{k'^2}\Delta P_{qq}^{(0)}(z)f_{NS}(\frac{x}{z},k'^2)
\end{equation}
The source of the double logarithmic terms $\ln^2x$ in $g_1(x,Q^2)$ is
the double integration in the formula for function $f(x,k^2)$:
\begin{equation}\label{r2.11}
f(x,k^2)\sim\frac{\alpha_s}{2\pi}\int\limits_x^1\frac{dz}{z}
\int\limits_{k_0^2}^{k^2/z}\frac{dk'^2}{k'^2}
\end{equation}
where the upper limit in the integral over the transverse momentum
$k'^2$ is $z$-dependent  ($=k^2/z$). Thus, double logarithmic terms
come from the integration over the longitudinal momentum fraction $z$
together with the integration over $k'^2$ with $z$-dependent upper
limit:
\begin{equation}\label{r2.12}
f(x,k^2)\sim\ln^2(1/x)=\ln^2x
\end{equation}
The analytical solution of (\ref{r2.10}) in the case of the fixed
coupling constant $\alpha_s$ \cite{b12} shows the singular small $x$
behaviour of the polarized, nonsinglet structure function $g_1$ {\it i.e.}:
\begin{equation}\label{r2.13}
g_1^{NS}(x,Q^2)\sim x^{-\lambda_{NS}};
\end{equation}
\begin{equation}\label{r2.14}
\lambda_{NS}=2\sqrt{\frac{\alpha_s}{2\pi}\Delta P_{qq}^{(0)}(x)};
\end{equation}
where $\Delta P_{qq}^{(0)}(x)$ is the splitting function and in the limit 
$x\rightarrow 0$ is equal to $4/3$. Hence for the fixed $\alpha_s$=0.18,
$\lambda_{NS}=0.39$ and as it has been already mentioned above, the singular
small $x$ shape of $g_1^{NS}$, implied by QCD dominates the REGGE behaviour:
\begin{equation}\label{r2.15}
REGGE:~~~~~g_1^{NS}(x,Q^2)\sim x^{-\alpha_{NS}(0)}~~~\alpha_{NS}(0)\le 0
\end{equation}
\begin{equation}\label{r2.16}
QCD:~~~~~g_1^{NS}(x,Q^2)\sim x^{-\lambda_{NS}}~~~\lambda_{NS}\sim 0.5
\end{equation}
In the case of the running coupling constant, numerical results for
$g_1^{NS}$, based on the Regge flat input parametrization
$g_1^{NS}(x,Q_0^2=1~{\rm GeV}^2)$ are given in \cite{b7}, \cite{b11}, \cite{b12}. 
In \cite{b13} the same method is used for dynamical input parametrization, 
where in contrast to Regge flat parametrizations, $g_1^{NS}(x,Q_0^2=1~{\rm GeV}^2)$ 
is singular for small $x$. The generation of the double logarithmic $\ln^2x$ 
terms is also possible via GLAP-like evolution equation for $g_1$ function 
with the rescaled transverse momentum squared. This alternative method is 
presented in the next section.

\section{GLAP-like evolution equation for $g_1^{NS}$, generating $\ln^2x$
terms at low $x$.}

We study the origin of the double logarithmic $\ln^2x$ terms in
$g_1^{NS}(x,Q^2)$ using a simple rescaling of $Q^2$ variable: 
$Q^2\rightarrow\mu^2=Q^2/x$, what leads to the GLAP-like equation for
$g_1^{NS}$ with evolution in a new scale $\mu^2$. We focus on the small $x$
region so our initial equation for further investigations is that which
contains only dominating $\ln^2x$ part. First let us consider the case with
a fixed coupling $\bar{\alpha_s}={\rm const}$, where
\begin{equation}\label{r3.1}
\bar{\alpha_s}=\frac{2\alpha_s}{3\pi}
\end{equation}
Thus the starting equation is:
\begin{equation}\label{r3.2}
f_{NS}(x,k^2)=f_{NS}^{(0)}(x,k^2)+\bar{\alpha_s}
\int\limits_x^1\frac{dz}{z}\int\limits_{k_0^2}^{k^2/z}
\frac{dk'^2}{k'^2}f_{NS}(\frac{x}{z},k'^2)
\end{equation}
After a simple substitution:
\begin{equation}\label{r3.3}
\mu^2=\frac{k^2}{x};~~~~x'=\frac{x}{z};~~~~\mu'^2=\frac{k'^2}{x'}
\end{equation}
the equation (\ref{r3.2}) takes a form:
\begin{equation}\label{r3.4}
f_{NS}(x,x\mu^2)=f_{NS}^{(0)}(x,x\mu^2)+\bar{\alpha_s}
\int\limits_x^1\frac{dz}{z}\int\limits_{k_0^2\frac{z}{x}}^{\mu^2}
\frac{d\mu'^2}{\mu'^2}f_{NS}(\frac{x}{z},\frac{x}{z}\mu'^2)
\end{equation}
After introduction an auxiliary function $\varphi(x,\mu^2)$:
\begin{equation}\label{r3.5}
\varphi(x,\mu^2)\equiv f_{NS}(x,x\mu^2)
\end{equation}
and applying Heaviside's $\Theta$ function:
\begin{equation}\label{r3.6}
\Theta(t)=\cases{1 & for~~ $t>0$ \cr 0 & for~~ $t\leq 0$ \cr}
\end{equation}
we get:
\begin{equation}\label{r3.7}
\varphi(x,\mu^2)=\varphi^{(0)}(x,\mu^2)+\bar{\alpha_s}
\int\limits_x^1\frac{dz}{z}\int\limits_{k_0^2}^{\mu^2}\frac{d\mu'^2}{\mu'^2}
\Theta(\mu'^2-k_0^2\frac{z}{x})\varphi(\frac{x}{z},\mu'^2)
\end{equation}
or
\begin{equation}\label{r3.8}
\varphi_{\Theta}(x,\mu^2)=\varphi_{\Theta}^{(0)}(x,\mu^2)+\bar{\alpha_s}
\int\limits_x^1\frac{dz}{z}\int\limits_{k_0^2}^{\mu^2}\frac{d\mu'^2}{\mu'^2}
\varphi_{\Theta}(\frac{x}{z},\mu'^2)
\end{equation}
where
\begin{equation}\label{r3.9}
\varphi_{\Theta}(x,\mu^2)\equiv
\Theta(\mu^2-\frac{k_0^2}{x})\varphi(x,\mu^2)
\end{equation}
The equation (\ref{r3.8}) has an exact form of the GLAP $Q^2$ evolution
formula for the unintegrated distribution function $f(x,Q^2)$. The
mentioned GLAP evolution equation for polarized nonsinglet quark
distributions $\Delta p$ (and hence for $g_1^{NS}$ function too) in the
small $x$ region is given by:
\begin{equation}\label{r3.10}
\frac{\partial\Delta p(x,Q^2)}{\partial\ln Q^2}=\bar{\alpha_s}
\int\limits_x^1\frac{dz}{z}\Delta p(\frac{x}{z},Q^2)
\end{equation}
Relation between $\Delta p(x,Q^2)$ and the unintegrated distribution
$f_p(x,Q^2)$ is as usual:
\begin{equation}\label{r3.11}
f_p(x,Q^2)=\frac{\partial\Delta p(x,Q^2)}{\partial\ln Q^2}
\end{equation}
what implies:
\begin{equation}\label{r3.12}
\Delta p(x,Q^2)=\Delta p_0(x)+
\int\limits_{Q_0^2}^{Q^2}\frac{dQ'^2}{Q'^2}f_p(x,Q'^2)
\end{equation}
where $\Delta p_0(x)$ is a nonperturbative part of $\Delta p$:
\begin{equation}\label{r3.13}
\Delta p_0(x)=\int\limits_{0}^{Q_0^2}\frac{dQ'^2}{Q'^2}f_p(x,Q'^2)
\end{equation}
and $Q_0^2=1~{\rm GeV}^2$ is the low scale of perturbative QCD. Hence the
evolution equation (\ref{r3.10}) written for the unintegrated distribution
function $f_p(x,Q^2)$ takes a form:
\begin{equation}\label{r3.14}
f_p(x,Q^2)=f_p^{(0)}(x,Q^2)+\bar{\alpha_s}
\int\limits_x^1\frac{dz}{z}\int\limits_{Q_0^2}^{Q^2}\frac{dQ'^2}{Q'^2}
f_p(\frac{x}{z},Q'^2)
\end{equation}
and
\begin{equation}\label{r3.15}
f_p^{(0)}(x,Q^2)=\bar{\alpha_s}\int\limits_x^1\frac{dz}{z}
\Delta p_0(\frac{x}{z})
\end{equation}
One can see from (\ref{r3.8}) and (\ref{r3.14}) that the double $\ln^2x$
resummation equation written for $\varphi_{\Theta}(x,\mu^2)$ function is a
GLAP $\mu^2$ evolution equation. The auxiliary function
$\varphi_{\Theta}(x,\mu^2)$ may be, similarly as in (\ref{r3.11}),
represented by an integrated function $u(x,\mu^2)$:
\begin{equation}\label{r3.16}
\varphi_{\Theta}(x,\mu^2)=
\frac{\partial u(x,\mu^2)}{\partial\ln\mu^2}
\end{equation}
and conversely:
\begin{equation}\label{r3.17}
u(x,\mu^2)=u_0(x)+\int\limits_{k_0^2}^{\mu^2}\frac{d\mu'^2}{\mu'^2}
\varphi_{\Theta}(x,\mu'^2)
\end{equation}
Thus the equation (\ref{r3.8}), generating double $\ln^2x$ effects can be
rewritten as:
\begin{equation}\label{r3.18}
\frac{\partial u(x,\mu^2)}{\partial\ln\mu^2}=\bar{\alpha_s}
\int\limits_x^1\frac{dz}{z} u(\frac{x}{z},\mu^2)
\end{equation}
Relation between the nonsinglet polarized structure function
$g_1^{NS}(x,Q^2)$ and the auxiliary function $u(x,\mu^2)$ is as follows:
\begin{equation}\label{r3.19}
g_1^{NS}(x,Q^2=x\mu^2)=u(x,\mu^2)
\end{equation}
In this way the problem of producing the $\ln^2x$ terms for
$g_1^{NS}(x,Q^2)$ in the small $x$ region via equation (\ref{r2.10}) has
been reduced to the GLAP evolution to the momentum scale $\mu^2=Q^2/x$. It
is not astonishing: appearing of the new evolution scale $Q^2/x$ has its
origin in the upper limit $k^2/z$ of the integration over the transverse
momentum in (\ref{r2.10}). This logarithmic integration over the transverse
momentum up to the $z-$dependent limit $k^2/z$ together with the logarithmic
integration over the longitudinal momentum fraction $z$ give double
logarithmic $ln^2x$ terms. The mechanism of appearing of the $ln^2x$ effects
in $g_1^{NS}$ from GLAP-like equations (\ref{r3.18})-(\ref{r3.19}) is well
visible just in a case of the fixed coupling constant
$\bar{\alpha_s}={\rm const}$. Then the eq.(\ref{r3.18}) can be solved
analytically. Using standard Mellin's method one can get the solution of 
eq.(\ref{r3.18}) in the form (see Appendix A):
\begin{equation}\label{r3.20}
u(x,\mu^2)\sim\sum\limits_{k=0}^{\infty}\frac{(\bar{\alpha_s}\ln\frac{1}{x}
ln\frac{\mu^2}{k_0^2})^k}{k!k!}
\end{equation}
and hence:
\begin{equation}\label{r3.21}
g_1^{NS}(x,Q^2)=u(x,\frac{Q^2}{x})\sim\sum\limits_{k=0}^{\infty}\frac
{[\bar{\alpha_s}\ln\frac{1}{x}(\ln\frac{1}{x}+\ln\frac{Q^2}{k_0^2})]^k}{k!k!}
\end{equation}
what gives approximately the leading small $x$ behaviour:
\begin{equation}\label{r3.22}
g_1^{NS}(x,Q^2)\sim x^{-2\sqrt{\bar{\alpha_s}}}
\end{equation}
Taking into account parton interactions through the introduction of the
running coupling constant one can get from (\ref{r3.18}) an equation which
incorporates the running couplings effects
$\bar{\alpha_s}\rightarrow\bar{\alpha_s}(Q^2)$:
\begin{equation}\label{r3.23}
\frac{\partial u(x,\mu^2)}{\partial\ln\mu^2}=\bar{\alpha_s}(x\mu^2)
\int\limits_x^1\frac{dz}{z} u(\frac{x}{z},\mu^2)
\end{equation}
However more justified theoretically seems the introduction of the running
coupling by the substitution $\bar{\alpha_s}\rightarrow\bar{\alpha_s}(Q^2/z)$, 
what gives:
\begin{equation}\label{r3.24}
\frac{\partial u(x,\mu^2)}{\partial\ln\mu^2}=\int\limits_x^1\frac{dz}{z}
\bar{\alpha_s}(\frac{x\mu^2}{z})u(\frac{x}{z},\mu^2)
\end{equation}
Our numerical analysis presented in the next section contains the both above
"running coupling" prescriptions and the constant $\bar{\alpha_s}$ case as
well.

\section{Numerical predictions for $g_1^{NS}(x,Q^2)$ based on the GLAP-like
equation, resumming the $\ln^2x$ terms.}

We solve numerically equations (\ref{r3.18}), (\ref{r3.23}) and (\ref{r3.24})
which are GLAP $\mu^2$ evolution equations for the auxiliary function
$u(x,\mu^2)$. The relation between $u(x,\mu^2)$ and the physical polarized
nonsinglet structure function $g_1^{NS}(x,Q^2)$ (\ref{r2.7}) is given by 
(\ref{r3.19}). In this way one can get the small $x$ behaviour of $g_1^{NS}$
governed by the double logarithmic $\ln^2x$ effects. Our predictions we
compare with those, received in the unintegrated $f(x,Q^2)$ approach
(\ref{r2.10}) and described in \cite{b7}. Solving the GLAP equations (\ref{r3.18}), 
(\ref{r3.23}) and (\ref{r3.24}) one should have an input parametrization of the 
$u$ function at the low scale $k_0^2$. Because $u(x,Q^2)$ has a meaning of the
physical $g_1^{NS}$ function for the rescaled $Q^2$ variable as it is shown
in (\ref{r3.19}), one can also write:
\begin{equation}\label{r4.1}
u(x,k_0^2)=g_1^{NS}(x,xk_0^2)
\end{equation}
The low scale $k_0^2$ introduced in (\ref{r3.7}) is equal to the usually
used QCD cut-off parameter $k_0^2=1~{\rm GeV}^2$. The really dependence on $z/x$
of the lower limit in the integration (\ref{r3.4}):
\begin{equation}\label{r4.2}
Q_0^2=\frac{z}{x}k_0^2
\end{equation}
became "shifted" to the definition of $\varphi_{\Theta}(x,\mu^2)$ via 
(\ref{r3.9}). It means that for running coupling cases one should take into
account in the evolution equations (\ref{r3.23}) and (\ref{r3.24}) the
cut-off factor $\Theta(\mu^2-k_0^2/x)$. Otherwise, the running variable
$x\mu^2$ in the coupling $\alpha_s$ becomes less than the low scale of QCD
evolution $k_0^2=1~{\rm GeV}^2$ what is definitely incorrect. There is no such a
constraint in the fixed coupling constant. Taking into account that for
perturbative QCD analysis the low cut-off parameter $k_0^2=1~{\rm GeV}^2$ we put
this value for the input scale of the $u$ function too. This implies that
for small $x$ the input parametrizations $u(x,k_0^2)$ corresponds to the 
$g_1^{NS}$ at the very low scale $xk_0^2$. We assume, that below the 
$k_0^2=1~{\rm GeV}^2$ the behaviour of the quark distributions is the same as at
$k_0^2$. Therefore we apply the standard parametrizations of the valence
quarks (and hence of $g_1^{NS}$) for the auxiliary function $u$:
\begin{equation}\label{r4.3}
u(x,k_0^2)=g_1^{NS}(x,k_0^2);~~~~~k_0^2=1~{\rm GeV}^2
\end{equation}
There are two basic kinds of input parametrizations of $g_1^{NS}(x,k_0^2)$:
the Regge one, which is flat at small $x$ and the singular one, which
behaves like $x^{-a} (a\sim 0.3)$ at small $x$. In our numerical
calculations we use two different input parametrizations: the Regge one,
which is given by
\begin{equation}\label{r4.4}
REGGE:~~~u(x,k_0^2)=\frac23g_A(1-x)^3=0.838(1-x)^3
\end{equation}
where $g_A=1.257$ is the axial vector coupling and the dynamical input GRSV
\cite{b15}:
\enlargethispage{0.2cm}
\begin{eqnarray}\label{r4.5}
GRSV&:&u(x,k_0^2)=0.327x^{-0.267}\nonumber\\
&\times&(1-0.583x^{0.175}+1.723x+3.436x^{3/2})(1-x)^{3.486}\nonumber\\
&+&0.027x^{-0.624}(1+1.195x^{0.529}+6.164x+2.726x^{3/2})(1-x)^{4.215}
\nonumber\\
\end{eqnarray}
For details about these parametrizations see also \cite{b13}. In all
calculations $\Lambda_{QCD} =232~{\rm MeV}$. Our numerical results are presented in
Figs.2-5. In Fig.2 the predictions for $g_1^{NS}$ at small $x$, based on the
eq. (\ref{r3.18}) for Regge and GRSV inputs respectively are shown. We use
two different value of fixed coupling: $\alpha_s=0.18$ and $\alpha_s=0.12$. The
input parametrizations are also plotted. In Fig.3 we confront the fixed
coupling results of eq.(\ref{r3.18}) for $\alpha_s=0.18$ with those, based
on eqs. (\ref{r3.23})-(\ref{r3.24}), taking into account running coupling
effects. Two groups of lines correspond to different inputs: Regge and GRSV.
The effective slope $\lambda_{NS}$ determined from (\ref{r2.16}) as:
\begin{equation}\label{r4.6}
\lambda_{NS}\sim\frac{\partial\ln g_1^{NS}}{\partial\ln\frac{1}{x}}
\end{equation}
for both input parametrizations and all mentioned above $\alpha_s$ cases:
$\alpha_s={\rm const}=0.18$, $\alpha_s(Q^2)$, $\alpha_s(Q^2/z)$ is presented in
Fig.4. The comparison of small $x$ predictions for $g_1^{NS}$ based on the
GLAP-like $\ln^2x$ approach with those based on the unintegrated function
$f(x,Q^2)$ and the eq.(\ref{r2.10}) is shown in Fig.5. Both parametrizations
are used. The running coupling effects $\alpha_s(Q^2)$ are included. We also
plot few small $x$ recent SMC 1997 data \cite{b4}. In all plots $Q^2=10~{\rm GeV}^2$.


\begin{figure}[hp]
\begin{center}
\includegraphics[width=90mm]{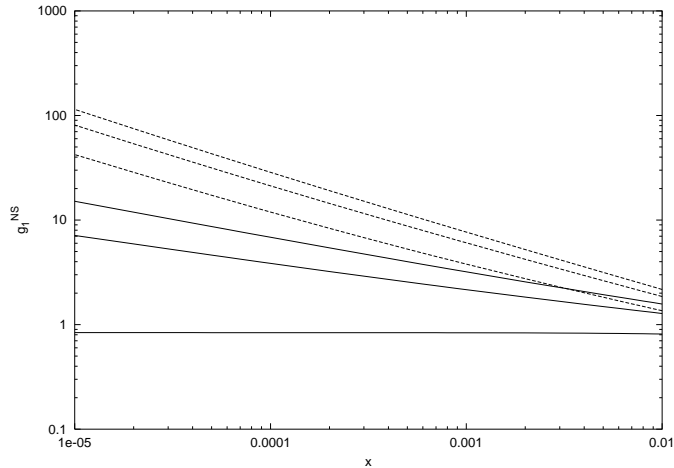}
\caption{The nonsinglet spin structure function of the proton 
$g_1^{NS}(x,Q^2)$ received from (\ref{r3.18}). Solid lines correspond to
Regge input, dashed lines - to GRSV input. In each group of lines (Regge or
GRSV) the lowest (for small $x$) line is the input at $k_0^2=1~{\rm GeV}^2$, 
the middle one concerns the fixed coupling $\alpha_s=0.12$ and the upper one 
is for $\alpha_s=0.18$. Evolution scale $Q^2=10~{\rm GeV}^2$.}
\end{center}
\end{figure}
\clearpage

\begin{figure}[ht]
\begin{center}
\includegraphics[width=85mm]{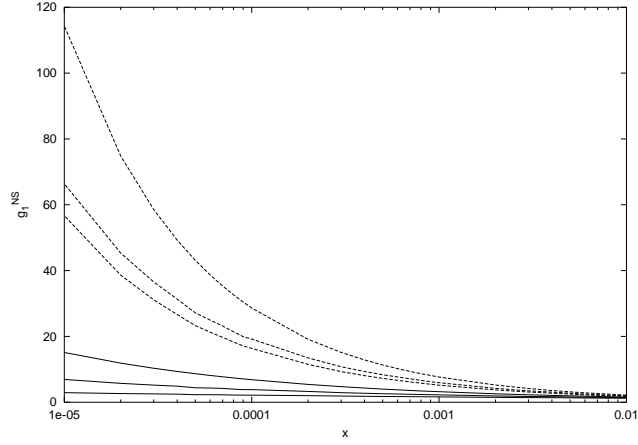}
\caption{The nonsinglet spin structure function of the proton 
$g_1^{NS}(x,Q^2=10~{\rm GeV}^2)$. Solid lines correspond to Regge input, 
dashed lines - to GRSV input. In each group of plots (Regge or GRSV) the
lowest line concerns the running coupling $\alpha_s(Q^2/z)$ case
(\ref{r3.24}), the middle one corresponds to $\alpha_s(Q^2)$ case
(\ref{r3.23}) while the upper plot is for fixed coupling
$\alpha_s={\rm const}=0.18$ (\ref{r3.18}).}
\end{center}
\end{figure}

\begin{figure}[hb]
\begin{center}
\includegraphics[width=85mm]{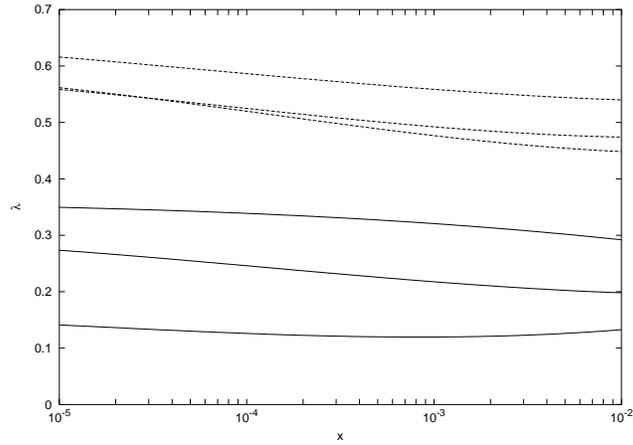}
\caption{Slope $\lambda_{NS}$ defined in (\ref{r4.6}) for different inputs
(Regge - solid, GRSV - dashed) and different $\alpha_s$. For each group of
lines (Regge or GRSV) the upper plot corresponds to fixed $\alpha_s=0.18$,
in the middle lies line for the running $\alpha_s(Q^2)$, and the lowest plot
is for the "more running" case: $\alpha_s(Q^2/z)$. $Q^2=10~{\rm GeV}^2$.}
\end{center}
\end{figure}
\clearpage

\begin{figure}[ht]
\begin{center}
\includegraphics[width=90mm]{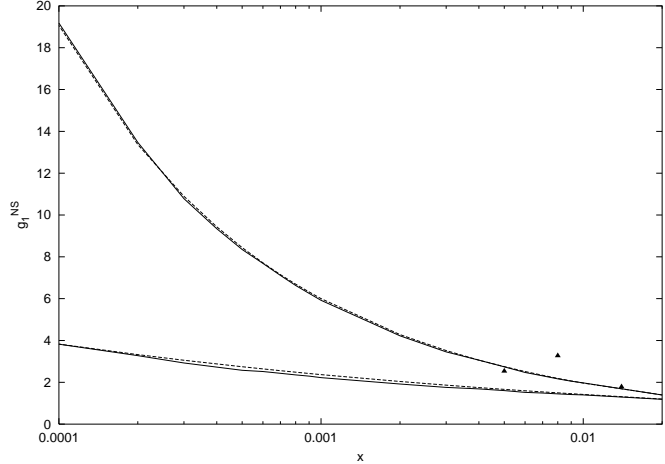}
\caption{Comparison of $g_1^{NS}$ predictions based on GLAP-like eq.
(\ref{r3.23}) - solid with that based on $f(x,Q^2)$ approach (\ref{r2.10}) -
dashed at $Q^2=10~{\rm GeV}^2$. Triangles show the recent small $x$ SMC
data 1997 \cite{b4}.}
\end{center}
\end{figure}


From Figs.2-5 one can read the following conclusions:

1.The $\ln^2x$ resummation gives steep growth of the structure function in
the small $x$ region. It is well known effect \cite{b6}, \cite{b7} that for 
$x\leq 10^{-2}$ the $\ln^2x$ terms dominate over the LO (or even NLO) 
evolution.

2.The growth of the structure function in the small $x$ region is of course
much steeper for the dynamical parametrization than for the flat one.
Singular inputs at $k_0^2$ intensify the QCD evolution effects while in the
case of flat parametrizations the singular small $x$ behaviour of structure
functions is completely and only generated by QCD evolution.

3.The effective slope $\lambda_{NS}$ (\ref{r4.6}) is contained between
0.14 for the nonsingular input and the running coupling $\alpha_s(Q^2/z)$
and 0.6 for the singular input and the fixed coupling $\alpha_s=0.18$. This
shows apart from the above conclusion in the point 2 that for the running
coupling constant $\alpha_s(Q^2)$, $\lambda_{NS}$ is smaller than for the
fixed one. Besides, introduction of the $\alpha_s=\alpha_s(Q^2/z)$ causes
the slope $\lambda_{NS}$ still smaller:
\begin{equation}\label{r4.7}
\lambda_{NS}(\alpha_s=const=0.18)>\lambda_{NS}(\alpha_s=\alpha_s(Q^2))>
\lambda_{NS}(\alpha_s=\alpha_s(Q^2/z))
\end{equation}
For comparison, the Regge theory predicts in the small $x$ region
$\alpha_{A_1}(0)\leq 0$ and $g_1^{NS}$ behaves as $x^{-\alpha_{A_1}(0)}\sim
{\rm const}$.

4.Comparing our results with those, based on the unintegrated function
$f(x,Q^2)$ and the equation (\ref{r2.10}) one can see agreement of these two
approaches. It is not astonishing because the both methods concern in
fact the same problem: resummation of the double logarithmic terms $ln^2x$
for the $g_1^{NS}$ function in the small $x$ region. Our approach based on
the GLAP-like equation (\ref{r3.18}) enables to keep the constance upper
scale of evolution $\mu^2$ instead of $Q^2/z$ like in (\ref{r2.10}). In this
way this approach become similar to the standard GLAP $Q^2$ evolution and we
can adopt the methods used in it to get new - $\ln^2x$ effects.

5.SMC recent data \cite{b4} for very small $x$ (3 points shown in Fig.5)
confirm the growth of the polarized nonsinglet structure function $g_1^{NS}$
in the small $x$ region. It seems that the singular inputs enable the better
agreement of theoretical predictions with experimental data.

\section{Summary and conclusions}

In this paper we have presented the double logarithmic $\ln^2x$ terms
resummation for the spin dependent nonsinglet structure function $g_1^{NS}$.
In our approach we have solved the GLAP-like equation for the auxiliary function
$u(x,\mu^2)$ which corresponds to the physical function
$g_1^{NS}(x,Q^2=x\mu^2)$ at the rescaled $\mu^2$ variable. Our calculations
have been performed for the simple nonsingular Regge parametrization and for
the dynamical one as well. Apart from the fixed coupling constant case, the
running coupling effects have been taken into account. Besides the effective
slope, controlling the small $x$ behaviour of $g_1^{NS}$ has been estimated.
Its value has been found lies between 0.14 for the nonsingular input and the 
running coupling $\alpha_s(Q^2/z)$ and 0.6 for the singular input and the fixed 
coupling $\alpha_s=0.18$. We found that the $\ln^2x$ effects govern the
small $x$ increase of the structure function $g_1^{NS}(x,Q^2)$. This growth
is larger in the case of the fixed coupling constant than for the running
one and of course for the singular input $g_1^{NS}(x,k_0^2)$ than for the
flat (e.g.Regge) one. The equation we have considered is not applicable for
the large $x$ ($x>10^{-2}$) region. Our formalism is however correct in the
very interesting small $x$ region. Presented results confirm that the
$\ln^2x$ effects are very significant for $x\leq 10^{-2}$. The spin
dependent structure functions of the nucleon are a sensitive test of the
perturbative QCD analyses in the small $x$ region. However practically lack
of experimental data in the very low $x$ region ($x\leq 10^{-3}$) causes the
satisfactory verification of the theoretical QCD predictions in this region
impossible. Also the predictions incorporating the double logarithmic
$\ln^2x$ effects in $g_1^{NS}$ are still awaiting for their crucial probe.

\section*{Acknowledgements}

We are grateful to Jan Kwieci\'nski for help in preparing this paper.

\appendix
\section{Analytical solution of the GLAP-like evolution equation for
$g_1^{NS}$ generating double logarithmic $\ln^2x$ effects at small $x$.}

We solve the equation (\ref{r3.18}) with the fixed coupling
$\bar{\alpha_s}=0.038$ using the standard Mellin method. The Mellin
transformation defines the $n-$ moment of $u(x,\mu^2)$ function by:
\begin{equation}\label{rA.1}
u^n(\mu^2)\equiv\int\limits_0^1 x^{n-1} u(x,\mu^2) dx
\end{equation}
Thus in moment space the evolution equation for small $x$ (\ref{r3.18}) is
simply given by:
\begin{equation}\label{rA.2}
\frac{du^n(\mu^2)}{d\ln\mu^2}= \frac{\bar{\alpha_s}}{n} u^n(\mu^2)
\end{equation}
The solution of (\ref{rA.2}) is straightforward:
\begin{equation}\label{rA.3}
u^n(\mu^2)=u_0^n \Bigl[\frac{\mu^2}{k_0^2}\Bigr]^{\frac{\bar{\alpha_s}}{n}}
\end{equation}
where $u_0^n$ is the $n-$ moment of the input $u(x,k_0^2)$ function:
\begin{equation}\label{rA.4}
u_0^n=\int\limits_0^1 x^{n-1} u(x,k_0^2) dx
\end{equation}
For the nonsingular input parametrization
\begin{equation}\label{rA.5}
u(x,k_0^2)\sim const
\end{equation}
$u_0^n$ has a form:
\begin{equation}\label{rA.6}
u_0^n = \frac{const}{n}
\end{equation}
Employing the inverse Mellin transformation:
\begin{equation}\label{rA.7}
u(x,\mu^2)=\frac{1}{2\pi i}\int\limits_{c-i\infty}^{c+i\infty} x^{-n}
u^n(\mu^2) dn
\end{equation}
which gives
\begin{equation}\label{rA.8}
u(x,\mu^2)=\frac{const}{2\pi i}\int\limits_{c-i\infty}^{c+i\infty}
\frac{x^{-n}}{n} \Bigl[\frac{\mu^2}{k_0^2}\Bigr]^{\frac{\bar{\alpha_s}}{n}} dn
\end{equation}
One can get the solution $u(x,\mu^2)$ of the form:
\begin{equation}\label{rA.9}
u(x,\mu^2)= const \sum\limits_{k=0}^{\infty}\frac{(\bar{\alpha_s}\ln\frac{1}{x}
ln\frac{\mu^2}{k_0^2})^k}{k!k!}
\end{equation}
Going back to the function $g_1^{NS}(x,Q^2=x\mu^2)$, one can find from 
(\ref{rA.9}) through (\ref{r3.19}) the following expression:
\begin{equation}\label{rA.10}
g_1^{NS}(x,Q^2)\equiv u(x,\frac{Q^2}{x})= const~{\it J_0}(2\sqrt
{\bar{\alpha_s}\ln\frac{1}{x}ln\frac{Q^2}{xk_0^2}})
\end{equation}
where ${\it J_0}(y)$ denotes modified Bessel function:
\begin{equation}\label{rA.11}
{\it J_0}(y) = \sum\limits_{k=0}^{\infty}\frac{(\frac{y}{2})^{2k}}{k!k!}
\end{equation}
This $\ln^2x$ terms resummation (\ref{rA.10}) gives in the small $x$ region
the effective behaviour of $g_1^{NS}$:
\begin{equation}\label{rA.12}
g_1^{NS}(x,Q^2) = x^{-\lambda_{NS}}
\end{equation}
where 
\begin{equation}\label{rA.13}
\lambda_{NS}\sim 2\sqrt{\bar{\alpha_s}}
\end{equation}
For the fixed $\bar{\alpha_s}=0.038$ the effective slope $\lambda_{NS}$ is
equal to 0.39.

\end{document}